\documentclass[12pt]{article}
\usepackage{amsfonts,amsmath,amssymb}
\usepackage{graphicx}
\usepackage{epstopdf}
\usepackage{todonotes}
\DeclareGraphicsExtensions{.eps}

\newcommand\encadremath[1]{\vbox{\hrule\hbox{\vrule\kern8pt
\vbox{\kern8pt \hbox{$\displaystyle #1$}\kern8pt}
\kern8pt\vrule}\hrule}}
\def\enca#1{\vbox{\hrule\hbox{
\vrule\kern8pt\vbox{\kern8pt \hbox{$\displaystyle #1$}
\kern8pt} \kern8pt\vrule}\hrule}}

\usepackage{tikz,mathtools}
\usetikzlibrary{knots}
\usetikzlibrary{calc}
\usetikzlibrary{decorations.pathreplacing} 
\usetikzlibrary{decorations.markings}
\usetikzlibrary{arrows}
\tikzset{->-/.style={decoration={
  markings,
  mark=at position .5 with {\arrow{>}}},postaction={decorate}}}

\makeatletter
\@addtoreset{equation}{section}
\makeatother
\newtheorem{theorem}{Theorem}[section]

\newtheorem{remark}{Remark}[section]
\newtheorem{proposition}{Proposition}[section]
\newtheorem{lemma}{Lemma}[section]
\newtheorem{corollary}{Corollary}[section]
\newtheorem{definition}{Definition}[section]

\newtheorem{example}[theorem]{Example}

\def\br{\begin{remark}\rm\small}
\def\er{\end{remark}}
\def\bt{\begin{theorem}}
\def\et{\end{theorem}}
\def\bd{\begin{definition}}
\def\ed{\end{definition}}
\def\bp{\begin{proposition}}
\def\ep{\end{proposition}}
\def\bl{\begin{lemma}}
\def\el{\end{lemma}}
\def\bc{\begin{corollary}}
\def\ec{\end{corollary}}
\def\bex{\begin{example}}
\def\eex{\end{example}}

\def\beaq{\begin{eqnarray}}
\def\eeaq{\end{eqnarray}}
\newcommand{\proof}{{\noindent \bf proof:}$\quad$ }
\newcommand{\eproof}{ $\square$ }

\usepackage[style=alphabetic,maxalphanames=4,giveninits,sorting=nyt,%
maxbibnames=99,backend=bibtex]{biblatex}
\renewbibmacro{in:}{}
\usepackage{csquotes}

\newcommand{\be}{\begin{equation}}
\newcommand{\beq}{\begin{equation}}
\newcommand{\eeq}{\end{equation}}
\newcommand{\bea}{\begin{eqnarray}}
\newcommand{\eea}{\end{eqnarray}}

\newcommand{\Tr}{{\operatorname{Tr}}}

\newcommand{\ii}{{\rm i}\,}

\newcommand{\CC}{{\mathbb C}}
\newcommand{\RR}{{\mathbb R}}
\newcommand{\ZZ}{{\mathbb Z}}

\newcommand{\spcurve}{{\cal S}}
\newcommand{\curve}{{\Sigma}}

\newcommand{\acycle}{{\cal A}}
\newcommand{\bcycle}{{\cal B}}

\newcommand{\xo}{x_{\circ}}
\newcommand{\zo}{z_{\circ}}

\newcommand{\Ker}{{\rm Ker\ }}

\newcommand{\formal}{\operatorname{formal}}
\newcommand{\GUE}{\operatorname{GUE}}

\newcommand{\diag}{\operatorname{diag}}

\newcommand{\x}{{\rm x}}
\newcommand{\y}{{\rm y}}
\newcommand{\F}{{\mathcal F}}

\newcommand{\td}{\tilde}
\newcommand{\Res}{\mathop{\,\rm Res\,}}
\usepackage[pdftex]{hyperref}
\hypersetup{colorlinks, urlcolor=magenta, citecolor=red, linkcolor=blue, filecolor=black}

\newcommand{\xstar}{x_*}
\newif\ifincludeproposal
\includeproposalfalse % true adds the separate, conditional reconstruction proposal
\begin{document}

\sloppy

\pagestyle{empty}
%\hfill IPhT-T17/086, CRM-3360
\addtolength{\baselineskip}{0.20\baselineskip}
\begin{center}
\vspace{26pt}
{\large \bf {Cylinder spectral curve and 1-cut Matrix model}}
\vspace{26pt}

{\sl B.\ Eynard}\footnote{Université Paris-Saclay, CNRS, CEA, Institut de physique théorique, 91191, Gif-sur-Yvette, France.}
\footnote{CRM, Centre de recherches math\'ematiques  de Montr\'eal, Montr\'eal, QC, Canada.}

\end{center}
\vspace{20pt}
\begin{center}
{\bf Abstract}
For a random matrix, with eigenvalues in a one-body potential $V(x)$ and a 2-body interaction $R(x,x')$, it is known that the resolvent converges in the large $N$ limit to an analytic function in the complex plane minus cuts, called the spectral curve. 
In the 1-cut case, a two-sheeted neighborhood of the cut is a cylinder with two simple ramification points.  
We describe these spectral curve data by boundary integral operators and generalized cycles.
Vice-versa, given a 2-branchpoints spectral curve in a cylinder, there exists a formal random matrix model that corresponds to it.
The goal of this article is to show the method to recover a formal random matrix model, with its potential and 2-body interaction, from the geometry of the spectral curve.

\end{center}

%-----------------------------ABSTRACT--------------------------------------
%
%Abstract

%\vspace{0.5cm}
%\section*{}

%\newpage
\vspace{26pt}
\pagestyle{plain}
\setcounter{page}{1}

%****************************************************************
%==================== ARTICLE =======================================%*******************
%*********************************************

\tableofcontents

%================================================
%================================================

\section{Introduction}

%================================================
%================================================

\medskip

Topological Recursion was first invented in order to compute the large size expansion of eigenvalues correlation function of a random matrix. 
The large $N$ eigenvalues density converges to a distribution, whose resolvent (Stieltjes transform) is an analytic function on the complex plane with cuts on the support of the limitting eigenvalues distribution. The Riemann surface (plane with cuts) and the resolvent, as well as the large $N$ limit of the 2-point correlation function constitute the spectral curve.

Then in \cite{EO07} the topological recursion formalism was extended beyond matrix model: from any spectral curve (a Riemann surface with an analytic $1$-form and $1\boxtimes 1$ form on it), one can recursively associate a double sequence of differential forms $\omega_{g,n}$.
In the matrix case, the $\omega_{g,n}$ are the coefficients of the asymptotic expansion of the $n$-point correlation functions.

\smallskip 
{\it Is every spectral curve  the spectral curve of a matrix model ?}

In general not, but here we show that if the surface is a cylinder with 2 branch-points, with positive holonomy around the cylinder, the answer is almost yes: there is indeed a matrix model of which it is the spectral curve.
%It is such that the $\omega_{g,n}$ are the coefficients of the large $N$ expansion of the Stieltjes transforms of $n$-point eigenvalues correlation functions.

% AI proposal
%\medskip
%
%Topological recursion first arose in the computation of the large-size expansion of matrix-model correlation functions. In a regular convergent regime, the planar resolvent is the Stieltjes transform of the equilibrium measure. In the formal setting, it is the leading coefficient of the topological expansion. Together with the planar connected two-point function, it determines the spectral data.
%
%In \cite{EO07}, this construction was extended beyond matrix models to spectral curves equipped with a meromorphic function $x$, a one-form, and a symmetric bidifferential, subject to the usual local regularity assumptions.
%
%\smallskip
%\emph{Which spectral curves arise from matrix models?}
%
%Here we study this question for a cylinder with two simple ramification points. The forward construction is described in Section~2. For the converse, the period $t>0$ is not by itself sufficient: one must also select a physical exterior, impose the corresponding asymptotic conditions on the bidifferential, and establish invertibility of the boundary response operator. Section~3 records the reconstruction problem and working calculations. Identification of all stable topological-recursion correlators with a matrix-model expansion requires the loop-equation and uniqueness hypotheses discussed in \cite{BEO13,BGK15}.
%

\subsection{Matrix integrals}

Let $N\geq 1$ be an integer, and $\hbar>0$ be a positive non-vanishing real number.
We consider a function $V(x)$, analytic in an open disc $D\subset \CC$, of radius large enough (which we shall make more precise below).
Let also $R(x,x')=R(x',x)$ be a symmetric bivariate function analytic in $D\times D$, and not singular on the diagonal.

We first use the following expression as a formal measure. 
\beq
\label{eq:defRMmeasure}
d\mu(\Lambda) = \frac{1}{N! Z_N(V,R)} \ 
d\Lambda_1\dots d\Lambda_N
\ \prod_{i<j} (\Lambda_i-\Lambda_j)^2 \prod_{i,j} e^{\frac12 R(\Lambda_i,\Lambda_j)} e^{- \hbar^{-1} \sum_i V(\Lambda_i)}.
\eeq
For a convergent probability measure one should additionally specify an  integration contour, reality, positivity, confinement and stability assumptions, see for example \cite{BGK15}.

\subsubsection{GUE}

GUE means Gaussian Unitary Ensemble, it is the case where the measure is Gaussian, more precisely it is the case
\bea
V(x) &=& \frac12 x^2 \cr 
R(x,x') &=& 0.
\eea
The integral of the measure on $\RR^N$ can be computed explicitly and equals
\bea
Z^{\text{GUE}}
&=& Z_N(V=\frac12 x^2,R=0) \cr 
&=& \frac{1}{N!}\int_{\RR^N}
d\Lambda_1\dots d\Lambda_N
\ \prod_{i<j} (\Lambda_i-\Lambda_j)^2 e^{- \hbar^{-1} \sum_i \frac12 \Lambda_i^2 } \cr 
&=& (2\pi)^{N/2}\hbar^{N^2/2} G(N+1).
\eea
where $G(N+1)=\prod_{j=0}^{N-1} j!$ is the Barnes function.

Using the 't Hooft parameter
\beq
\label{eq:deftHooftparameter}
t= \hbar N,
\eeq
and the Barnes asymptotic expansion, we obtain
\begin{align}
\ln Z^{\GUE}
&\sim N\ln(2\pi)+\frac{N^2}{2}\ln\hbar
 +\left(\frac{N^2}{2}-\frac1{12}\right)\ln N-\frac34N^2
 \nonumber\\
&\quad +\frac1{12}-\ln A
 +\sum_{g\geq2}\frac{B_{2g}}{2g(2g-2)}N^{2-2g}
 \nonumber\\
&=N\ln(2\pi)+\frac1{12}\ln\hbar
 +\sum_{g\geq0}\hbar^{2g-2}\F_g^{\GUE}(t).
\label{eq:Gaussian-normalization}
\end{align}

where $B_k$ is the $k^{th}$ Bernoulli number, and $A=1.282...$ is the Glaisher's constant, and where
\bea
\F^{\GUE}_0  &=& \frac12 t^2 \ln t - \frac34 t^2 \cr 
\F^{\GUE}_1  &=& -\frac{1}{12} \ln t +\frac{1}{12} - \ln A \cr 
\F^{\GUE}_g  &=& \frac{B_{2g}}{2g(2g-2)} t^{2-2g} \quad \text{for } g\geq 2.
\eea

\subsubsection{General case: formal matrix integrals}

Consider the integral
\bea
Z_N(V,R) 
%&=&  \frac{1}{N!}
%\int_{N\times N}^{\formal} d\mu(\Lambda) \cr 
&=& \frac{1}{N!}
\int_{N\times N}^{\formal}
d\Lambda_1\dots d\Lambda_N
\ \prod_{i<j} (\Lambda_i-\Lambda_j)^2 \prod_{i,j} e^{\frac12 R(\Lambda_i,\Lambda_j)} e^{- \hbar^{-1} \sum_i V(\Lambda_i)} \cr
&=& 
\frac{1}{N!} \int_{N\times N}^{\formal}
d\Lambda_1\dots d\Lambda_N
\  \prod_{i<j} (\Lambda_i-\Lambda_j)^2 e^{R(\Lambda_i,\Lambda_j)}  \cr 
&& \qquad   \prod_{i=1}^N e^{- \hbar^{-1} (V(\Lambda_i)-\frac12 \hbar R(\Lambda_i,\Lambda_i))} . 
\eea
over diagonal $N\times N$ matrices $\Lambda=\diag(\Lambda_1,\dots,\Lambda_N)$.
%This can also be written
%\bea
%N! Z_N(V,R) 
%&=& 
%\int_{N\times N}^{\formal}
%d\Lambda_1\dots d\Lambda_N
%\ \prod_{i<j} (\Lambda_i-\Lambda_j)^2 e^{R(\Lambda_i,\Lambda_j)}   \prod_{i=1}^N e^{- \hbar^{-1} (V(\Lambda_i)-\frac12 \hbar R(\Lambda_i,\Lambda_i))} . \cr
%\eea

``Formal integral'' (see \cite{EynBook}) means that one should first consider a critical point $\xstar$ of $V(x)$, i.e. a non-degenerate solution of $V'(\xstar)=0$ with $V''(\xstar)\ne0$, then rescale $\Lambda_i = \xstar + \hbar^{\frac12} \lambda_i$, and expand the integrand as a series in powers of $\hbar$.
For each power of $\hbar$, the integral is a polynomial moment of a Gaussian integral, and can thus be computed by Wick's theorem.
The formal integral is the formal series of $\hbar$ whose coefficients are these polynomial Gaussian moments:
\bea
\frac{Z_N(V,R)}{Z^{\text{GUE}}}
&=& e^{- \hbar^{-1} N  V(\xstar)} e^{\frac12 N^2 R(\xstar,\xstar)} 
V''(\xstar)^{-N^2/2}  \left( 1 + O(\hbar) \right).
\eea
This estimate is a fixed-$N$ saddle expansion; its remainder is not asserted to be uniform when $N=t/\hbar$. The fixed-$t$ topological expansion below is a separate formal reorganization.
The disc $D$ should be a disc containing $\xstar$ and no other critical points of $V$, and of radius $>0$.
% large enough compared to $t=N\hbar$.
Let us mention that the formal integral depends on the choice of critical point of $V$: different choices lead to different formal integrals. See for instance chapter 4 of \cite{EynBook}.

Instead of formal integrals as in \cite{EynBook}, we may also consider more generally the case where the integral of the measure generates an asymptotic series as powers of $\hbar$
For example in \cite{BGK15} some conditions are given so that an actual matrix integral has a large $N$ asymptotic expansion as a series in $\hbar$. 

\medskip

%We then define the formal random matrix measure
%\bea
%\label{eq:defRMmeasure}
%d\mu(\Lambda) &=& \frac{1}{N! Z_N(V,R)} \ 
%d\Lambda_1\dots d\Lambda_N
%\ \prod_{i<j} (\Lambda_i-\Lambda_j)^2 \prod_{i,j} e^{\frac12 R(\Lambda_i,\Lambda_j)} e^{- \hbar^{-1} \sum_i V(\Lambda_i)}.
%\eea

With the formal measure \eqref{eq:defRMmeasure} we define the resolvent correlators:
\bea
\hat W_n(x_1,\dots,x_n)
&=& \mathbf E\left( \prod_{i=1}^n \Tr \frac{1}{x_i-\Lambda} \right)
\eea
and their cumulants
\bea
W_n(x_1,\dots,x_n)
&=& \mathbf E_{\operatorname{cumulant}}\left( \prod_{i=1}^n \Tr \frac{1}{x_i-\Lambda} \right)
\eea
In the formal topological expansion at fixed $t=\hbar N$, we write:
\bea
W_n(x_1,\dots,x_n)
&=& \sum_{g=0}^\infty \hbar^{2g-2+n} W_{g,n}(x_1,\dots,x_n).
\eea
For $n=0$ we define
\bea
\ln \left(Z_N(V,R)/Z^{\text{GUE}}\right)
&=& \sum_{g=0}^\infty \hbar^{2g-2} (\F_g- \F^{\GUE}_g) .
\eea
Under the regularity, solvability and projection hypotheses for the loop equations, the coefficients are computed by topological recursion \cite{BEO13}. For convergent rather than formal integrals, existence of a full expansion requires additional analytic assumptions \cite{BGK15}.

\section{From matrix model to spectral curve}

\subsection{Roots of the 2-point interaction}

%
%Consider that $e^{R(x,x')}$ has some zeros at $x'=\sigma_j(x)$ of degree $\alpha_j>0$ or poles at $x'=\sigma_j(x)$ of degree $-\alpha_j>0$, labeled by some $j\in J$, so that we can write
%\bea
%e^{R(x,x')} &=& \prod_{j\in J} ((x'-\sigma_j(x))^{\alpha_j} \cr
%R(x,x') &=&  \sum_{j\in J} \alpha_j \ln{(x'-\sigma_j(x))} ,
%\eea
%and thus its derivative with respect to the first variable is
%\bea
%R_x(x,x') &=& \frac{\partial  R(x,x')}{\partial x} 
%= \sum_{j\in J} \alpha_j \frac{\sigma_j'(x)}{\sigma_j(x)-x'}.
%\eea
%If the product is an infinite product, we assume that it can be regularized by Zeta-regularization.
%
%The group generated by all the functions $x\mapsto \sigma_j(x)$ is the Galois group associated to $R(x,x')$.
%A famous example is the $O(n)$ model for which $R(x,x')=-n\ln{(x+x'+m)}$, so that $R_x(x,x') = \frac{-n}{x'+x+m}$ and thus $\sigma(x)=-m-x$ and $\alpha=-n$, and the Galois group  $G=\{\text{Id},\sigma\}$ is isomorphic to $\ZZ_2$.
%
%
%The symmetry $R(x,x') = R(x',x)$ implies that
%\bea
%R(x,x') &=&  \sum_{j\in J} \alpha_j \ln{(x'-\sigma_j(x))}  = \sum_{j\in J} \alpha_j \ln{(x-\sigma_j(x'))}  \cr
%\eea
%and thus
%\bea
%\label{eq:symRx}
%R_x(x,x') = \frac{\partial  R(x,x')}{\partial x} 
%&=& \sum_{j\in J} \alpha_j \frac{1}{x-\sigma_j(x')} \cr 
%&=& \sum_{j\in J} \alpha_j \frac{\sigma_j'(x)}{\sigma_j(x)-x'}.
%\eea
%
%

In many applications, the 2-body interaction $R$ is such that $e^{R(x,\xi)}$ has zeros and poles.
Let us call it the logarithmic hypothesis:
\begin{align}
\text{Log hypothesis : } \qquad R(x,\xi)&=\sum_{j\in J}\alpha_j\log(\xi-\sigma_j(x)) .
\end{align}
where, if $J$ is infinite, we assume that the sum can be defined by Zeta-regularization.  The finite-$J$ case avoids that additional issue.
Under that assumption we have 
\begin{align}
R(x,\xi)&=\sum_{j\in J}\alpha_j\log(\xi-\sigma_j(x)) 
\label{eq:root-ansatz}\\
R_x(x,\xi)&=\sum_{j\in J}\alpha_j
\frac{\sigma_j'(x)}{\sigma_j(x)-\xi} .
\end{align}

If the maps $\sigma_j$ are invertible on appropriate domains they generate a group or a local pseudogroup. 

A famous example is the $O(n)$ model for which $R(x,x')=-n\ln{(x+x'+m)}$, so that $R_x(x,x') = \frac{-n}{x'+x+m}$ and thus $\sigma(x)=-m-x$ and $\alpha=-n$, and the group $G=\{\text{Id},\sigma\}$ is isomorphic to $\ZZ_2$.

%; we do not call this a Galois group without specifying the associated covering.

%For the $O(n)$ model,
%\begin{equation}
%R(x,\xi)=-n\log(x+\xi+m),\qquad
%\sigma(x)=-m-x,\qquad \alpha=-n.
%\end{equation}
%Here $\sigma^2=\mathrm{Id}$ and the generated group is $\ZZ_2$.

\subsection{Resolvent}

We already  introduced the resolvent and its expansion in powers of $\hbar$ as
\bea
W(x) = W_1(x) =  \mathbb E \left( \sum_{i} \frac{1}{x-\lambda_i} \right)
 = \sum_{g=0}^\infty \hbar^{2g-1} W_{g,1}(x).
\eea
From its definition, the leading order $W_{0,1}$ should behave at $x\to\infty$ like
\beq
W_{0,1}(x) \mathop{\sim}_{x\to \infty} N\hbar \ \frac{1}{x} =  \frac{t}{x},
\eeq 
i.e. $W_{0,1}(x)dx$ is meromorphic in a neighborhood of $\infty$, with a simple pole at $\infty$ of residue $-t=-\hbar N$.
It can be analytically continued to the complex plane, with some cuts, and it is discontinuous across the cuts.

For convergent models satisfying suitable confinement, stability and off-criticality assumptions, the planar limit is the Stieltjes transform of an equilibrium measure; see \cite{DFGZJ,RMTBookEynKimRib,BGK15}. In the formal setting, the following saddle-point equation is instead understood coefficientwise. The energy whose stationarity gives it is
\bea
\hbar^{-1}\sum_iV(\Lambda_i)-\sum_{i<j}\log(\Lambda_i-\Lambda_j)^2
-\frac12\sum_{i,j}R(\Lambda_i,\Lambda_j).
\eea

Taking the derivative w.r.t. $\Lambda_i$ gives
%\bea
%\hbar^{-1} V'(\Lambda_i) 
%&=& 2\sum_{j\neq i} \frac{1}{\Lambda_i-\Lambda_j} + \sum_j R_x(\Lambda_i,\Lambda_j) \cr 
%&=& 2\sum_{j\neq i} \frac{1}{\Lambda_i-\Lambda_j} + \sum_j\sum_{k\in J} \alpha_k \frac{\sigma_k'(\Lambda_i)}{\sigma_k(\Lambda_i)-\Lambda_j} \quad \text{under log hypothesis}. \cr 
%\eea
\bea
\label{eq:dscrsdpteqW1}
\forall x=\Lambda_i \ : && \cr 
\hbar^{-1} V'(x) 
&=& 2\sum_{j\neq i} \frac{1}{x-\Lambda_j} + \sum_{j=1}^N R_x(x,\Lambda_j) \cr 
&=& 2\sum_{j\neq i} \frac{1}{x-\Lambda_j} + \sum_{j=1}^N\sum_{k\in J} \alpha_k \frac{\sigma_k'(x)}{\sigma_k(x)-\Lambda_j}  \quad \text{under log hypothesis}. \cr
\eea
In the $\hbar\to 0$ limit, the leading order $W_{0,1}(x) = \hbar\sum_j 1/(x-\Lambda_j)$ satisfies, along each of its cuts the famous "saddle point equation" \cite{DFGZJ,RMTBookEynKimRib,BEO13} which is formally the large $N$ limit of \eqref{eq:dscrsdpteqW1}
\bea
\label{eq:RMsdptequ}
V'(x) &=& W_{0,1}(x+\ii 0)+W_{0,1}(x-\ii 0) + \frac{1}{2\pi\ii} \oint_{\mathcal C_+} R_x(x,\xi) W_{0,1}(\xi) d\xi \cr 
V'(x) &=& W_{0,1}(x+\ii 0)+W_{0,1}(x-\ii 0) + \sum_{k\in J} \alpha_k \sigma'_k(x)  W_{0,1}(\sigma_k(x)) \cr 
&& \qquad \qquad \qquad \quad\quad\quad \quad \quad\quad \quad \text{under log hypothesis},
\eea
where $\mathcal C_+$ is a circle surrounding the cut, oriented with the cut on its left, and not enclosing any singularity of $R_x(x,\xi)$.
This equation is derived in many works on random matrices and potential theory. Its solutions are well studied.

There are many examples of $V$ and $R$, such that the solution $W_{0,1}$ is an algebraic function of $x$. 
Root-system and finite-group criteria for algebraicity are available in specific classes of interactions \cite{BEroots}; no such criterion will be assumed here, we just consider a solution, algebraic or not.
Our goal is to reformulate \eqref{eq:RMsdptequ} in a way more convenient for our purposes.

\subsection{One-cut case and the annulus around the cut}

We assume a regular soft-edge one-cut solution supported on an arc $[b,a]$ homotopic and close enough to the straight segment $[b,a]$. Its two-sheeted neighborhood is an annulus. 
%The Joukowsky map below sends the unit circle to a straight segment.

\subsubsection{Joukowsky parametrization of the annulus}

Let us introduce the Joukowsky parametrization:
\bea
\x(z) &=& \alpha + \beta (z+1/z) \cr 
\alpha &=& \frac{a+b}{2} \cr 
\beta &=& \frac{a-b}{4}
\eea 
or vice-versa $a=\alpha+2\beta$, $b = \alpha-2\beta$.
In the $z$ variable, the cut is a closed Jordan loop invariant under the involution $\mathfrak{i}(z) = \frac{1}{z}$, and is homotopic to the unit circle $|z|=1$:
\beq
\text{cut} = \x(\mathcal C_0 ) \qquad \text{ with } \mathcal C_0 \text{ homotopic to } S^1 \text{ in the annulus}. 
\eeq
A neighborhood of the cut is an annulus $1/r<|z|<r$ for some $r>1$.

We introduce the involution on the annulus
\bea
\mathfrak{i}(z) = \frac{1}{z} .
\eea
In addition to the one-cut hypothesis, assume that the interaction singularities are separated from the cut. After shrinking a collar if necessary, choose $r>1$ such that for all $1/r<|z|<r$, $R_x(\x(z),\x(z'))$ is analytic in the domain $1/r<|z'|<r$.
In the log-hypothesis, denoting $\td\sigma_j$ be the physical-sheet lift satisfying
$\x(\td\sigma_j(z))=\sigma_j(\x(z))$, 
 this implies that all singularities $\td \sigma_j(z)$ are outside of the annulus, i.e. for $1<|z|<r$:
\begin{equation}
|\td\sigma_j(z)|>r .
\label{eq:local-separation}
\end{equation}
%
%We assume analyticity in a slightly larger collar so that boundary values and contour deformations are defined. This is a local separation condition, not a condition on all $|z|>1$. For infinitely many maps it includes a uniform separation requirement. For the exterior residue proof we also assume $R_x(x,\xi)=O(\xi^{-1})$ as $\xi\to\infty$ and no additional exterior singularities beyond those represented in \eqref{eq:root-ansatz}.

\subsubsection{Some geometry of the cylinder}
\label{sec:geometryCylinder}

The annulus (or cylinder) is the Riemann surface 
\beq
\curve = \CC \cap \{1/r\leq |z| \leq r\} .
\eeq 
Its boundary is made of 2 circles, oriented such that the annulus is on their left:
\bea
\partial\curve &=& \mathcal C = \mathcal C_+ + \mathcal C_- \cr 
\mathcal C_+ &=&  \left\{r e^{\ii\theta} \right\}_{\theta\in [0,2\pi]} \cr 
\mathcal C_- &=&  \left\{\frac{1}{r} e^{-\ii\theta} \right\}_{\theta\in [0,2\pi]}  .
\eea
We define the 2 half-cylinders
\bea
\curve_+ &=& \CC \cap \{1< |z| \leq r\} \qquad \text{called upper sheet or physical sheet} \cr 
\curve_- &=& \CC \cap \{1/r \leq  |z| <1\} \qquad \text{called lower sheet } .
\eea

Let us also define the discs
\bea
\mathcal D_+ &=&  \CC P_1 \cap \{|z|>1\} \cr 
\mathcal D_- &=&  \CC P_1 \cap \{|z|<1\} . 
\eea
and
\bea
\curve &=& \CC P_1 \cap \{1/r\leq |z|\leq r\} \cr 
\curve \cup \mathcal D_+ &=& \CC P_1 \cap \{|z|\geq 1/r\} \cr 
\curve \cup \mathcal D_- &=& \CC P_1 \cap \{|z|\leq r\} \cr 
\curve_+ \cup \mathcal D_+ &=& \mathcal D_+ = \CC P_1 \cap \{|z|>1\} \cr 
\curve_- \cup \mathcal D_- &=& \mathcal D_- = \CC P_1 \cap \{|z|<1\} \cr 
\curve \cup \mathcal D_+ \cup \mathcal D_- &=& \CC P_1 .
\eea
In other words we may glue the discs $\mathcal D_\pm$ at the boundaries of the cylinder $\curve$ completing the cylinder to either a disc around $\infty$ or a disc around $0$ or the full $\CC P^1$.

The following lemma is well known and very useful
\bl[Canonical decomposition]
\label{Lem:Laurentdecomp}
Every function $f(z)$ analytic in the cylinder can be decomposed as 
\bea
f(z) = f_+(z)+f_-(z)
\eea 
where $f_+$ can be analytically continued to $\mathcal D_+$ and $f_-$ can be analytically continued to $\mathcal D_-$.
Namely:
\bea
f_-(z) &=& \sum_{k=0}^\infty f_k z^k \cr 
f_+(z) &=& \sum_{k=1}^\infty f_{-k} z^{-k} \cr 
f(z) &=& \sum_{k\in \ZZ} f_k z^k .
\eea
The fact that $f_+$ (resp. $f_-$) is analytic in $|z|>1/R$ (resp. $|z|<R$) implies
\beq
\label{eq:boundfk}
\forall 1<r<R \qquad
|f_k| = O(r^{-|k|}).
\eeq

Every 1-form $\omega(z)$ analytic in the cylinder can be uniquely decomposed as 
\bea
\omega = \omega_++\omega_-+\omega_0
\eea
where
\bea
\label{eq:defomegapm0}
\omega_0 &=& \frac{dz}{z} \ \left( \frac{1}{2\pi\ii} \oint_{S^1} \omega \right) \cr 
\omega_+ &=& \sum_{k\geq 1} \omega_{+,k} z^{k-1} dz \cr 
\omega_- &=& \sum_{k\geq 1} \omega_{-,k} z^{-k-1} dz .
\eea
The fact that $\omega_+$ (resp. $\omega_-$) is analytic in $|z|>1/R$ (resp. $|z|<R$) implies
\beq
\label{eq:boundomegak}
\forall 1<r<R \qquad
|\omega_{\pm,k}| = O(r^{-|k|}).
\eeq

%We define the projection operators $\Pi^0,\Pi^+,\Pi^-$ as
%\bea
%\label{eq:defPiprojections}
%\Pi^0.\omega = \omega_0
%\quad , \quad
%\Pi^+.\omega = \omega_+
%\quad , \quad
%\Pi^-.\omega = \omega_- .
%\eea

\el
\proof{}
For $z\in\curve$, define 
\bea 
f_-(z) &=& \frac{1}{2\pi\ii} \oint_{\mathcal C_+} f(z') \frac{dz'}{z'-z} = \frac{1}{2\pi\ii} \oint_{\mathcal C_+} f(z') \sum_{k=0}^\infty z^k \frac{dz'}{z'^{k+1}} \cr 
f_+(z) &=& \frac{1}{2\pi\ii} \oint_{\mathcal C_-} f(z') \frac{dz'}{z'-z} = -\frac{1}{2\pi\ii} \oint_{\mathcal C_-} f(z') \sum_{k=0}^\infty  \frac{z'^k dz'}{z^{k+1}} .
\eea
We clearly have $f_-(z)+f_+(z) = \Res_{z'\to z} f(z') dz'/(z'-z)=f(z)$.

For 1-forms, observe that $\omega-\omega_0$ has vanishing period around the cylinder, so it is an exact form that can be written as $\omega-\omega_0=df$. Then use the decomposition of $f$.
The bounds on the growth of coefficients at large $|k|$ is follows from  Cauchy-Hadamard.
\eproof

\bd
\label{Def:Pipm}
On the space $\Omega(\curve)$ of holomorphic 1-forms on $\curve$, define:
We also define, for $x\in \x(\curve^{(1)})$
\bea
\Pi^+ \omega(z)  
  &=&  \frac{d\x(z)}{2\pi\ii} \oint_{\mathcal C_+}  \frac{1}{\x(z)-\x(z')} \omega_+(z') \cr
&=&  \frac{d\x(z)}{2\pi\ii} \oint_{\mathcal C_+}  \frac{1}{\x(z)-\x(z')} (\omega(z')-\omega_0(z')) \cr
\Pi^- \omega(z)  
  &=&  \frac{d\x(z)}{2\pi\ii} \oint_{\mathcal C_-}  \frac{1}{\x(z)-\x(z')} \omega_-(z') \cr
&=&  \frac{d\x(z)}{2\pi\ii} \oint_{\mathcal C_-}  \frac{1}{\x(z)-\x(z')} (\omega(z')-\omega_0(z')) 
\eea

We also have
\bea
\label{eq:defPipm2}
\Pi^+.\omega &=& \sum_{k\geq 1} \frac{1}{k} \omega_{+,k} d(z^k+z^{-k}) \cr 
\Pi^-.\omega &=& -\sum_{k\geq 1} \frac{1}{k} \omega_{-,k} d(z^k+z^{-k}) .
\eea
which are absolutely convergent series in $\curve$ thanks to \eqref{eq:boundomegak}.
\ed

We have
\bl
\label{Lem:Pipmextend}
The 1-form
\beq
\omega^\pm = \omega - \Pi^\pm.\omega 
\eeq
is such that
$\omega^\pm-\omega_0$  can be holomorphically continued to $\curve_\pm\cup\mathcal D_\pm$. 
In other words, $\omega^\pm$ can be analytically continued to $\curve_\pm\cup\mathcal D_\pm$ with at most a simple pole at $z=\infty$ (resp. $z=0$).

Moreover:
\beq
\label{eq:Pioddsame}
(1-\mathfrak i)^*\omega^\pm = (1-\mathfrak i)^*\omega,
\eeq
i.e. $\omega$ and $\omega^\pm$ differ only by an even (under the involution) form, they have the same odd-part.

Moreover we have the expansion
\beq
\omega(z)-\omega_0(z) - \Pi^\pm\omega(z)
= \sum_{k\geq 1} \frac{d\x(z)}{(\x(z)-\xo)^{k}}  \frac{-1}{2\pi\ii} \oint_{\mathcal C_\pm} (\x(z')-\xo)^{k-1} \omega(z') . 
\eeq

\el
\proof{}
Let $t_0= \frac{1}{2\pi\ii} \oint_{S^1} \omega$.
If the lemma holds in the case $t_0=0$, then it holds for any $t_0$, so let us prove it in the case $t_0=0$.

$\Pi_\pm\omega$ is an absolutely convergent integral for any $z\in\curve$.

Then  consider $|z|>R$, the following integral
\bea
A_+(z)  &=&  \frac{d\x(z)}{2\pi\ii} \oint_{\mathcal C_+}  \frac{1}{\x(z)-\x(z')} \omega_+(z') 
\eea
is also an absolutely convergent integral, it defines an analytic 1-form on the domain $|z|>R$. 
It can be continued across $\mathcal C_+$, to the domain $|z|>1$, by deforming the integration contour, picking a residue at $z'=z$, therefore on $1<|z|<R$ we have
\bea
A_+(z)  &=&  - \omega_+(z) + \Pi^+\omega(z).
\eea
In other words $\omega_+ - \Pi^+\omega$ can be analytically continued to $\curve_+\cup\mathcal D_+$.
Moreover, since $\omega_-$ can also be analytically continued to $\curve_+\cup\mathcal D_+$, we have that 
$\omega - \omega_0 - \Pi^+\omega$ can be analytically continued to $\curve_+\cup\mathcal D_+$.
Same proof for $\omega - \omega_0 - \Pi^-\omega$ can be analytically continued to $\curve_-\cup\mathcal D_-$.

In equation~\eqref{eq:defomegapm0} $\omega_\pm$ is a holomorphic function in the annulus, therefore it is an absolutely convergent sum, and thus $|\omega_{\pm,k}| R^{k}=O(1)$. 
A fortiori this also implies that $|\omega_{\pm,k}| R^{-k}=O(1)$, and  thus each of equations \eqref{eq:defPipm2} is an absolutely converging sum, and defines an anlytic function in the annulus. It satisfies the same properties as $A_\pm$ and takes the same value at $\x(z)=\infty$, therefore they have to coincide.

Eqs \eqref{eq:Pioddsame} are satisfied because we subtracted an even quantity.
\eproof{}

%
%
%
%
%TODO HERE
%
%, we have
%\bea
%\omega_0 &=& t_0 \frac{dz}{z} \cr
%\omega &=& t_0 \frac{dz}{z} + \sum_{k=1}^\infty \frac{1}{k}\omega_{+,k} dz^{k} - \sum_{k=1}^\infty \frac{1}{k} \omega_{-,k} d z^{-k} 
%\eea
%therefore
%\bea
%\omega^+ &=& t_0 \frac{dz}{z} - \sum_{k=1}^\infty \frac{1}{k}\omega_{+,k} d z^{-k} - \sum_{k=1}^\infty \frac{1}{k} \omega_{-,k} d z^{-k} \cr 
%\omega^- &=& t_0 \frac{dz}{z} + \sum_{k=1}^\infty \frac{1}{k}\omega_{+,k} d z^{k} + \sum_{k=1}^\infty \frac{1}{k} \omega_{-,k} d z^{k} .
%\eea
%$\omega^+-\omega_0$ (resp. $\omega^--\omega_0$) contains only negative (resp. positive) powers of $z$, so it can be extended as a 1-form holomorphic in $\mathcal D_\pm$.
%
%Since the subtraction is even, it is obvious that $\omega^\pm$ and $\omega$ have the same odd part.
%\eproof

\subsubsection{Reference point}
\label{sec:referencePoint}

We also define a reference point $\xo$, as a generic point in $\x(\curve)$, and $\zo$ a chosen preimage in the closed upper collar $|\zo|\geq1$, away from branch points.
A convenient choice can be $\xo=\alpha$ and $\zo=\ii$, but in fact it is better to choose a reference point independent of the moduli of the spectral curve, and thus independent of $\alpha$ and $\beta$. 
A Choice often made is $\xo=0$, if the origin is indeed in $\x(\curve)$.

\subsubsection{Cauchy kernel and relations}

We define the fundamental 2nd kind form of the annulus
\beq
\td B(z,z') = \frac{dz \otimes dz'}{(z-z')^2}.
\eeq
This is the the restriction to the annulus, of the unique symmetric $1\boxtimes 1$ bidifferential on $\CC P^1\times \CC P^1$ that has a double pole on the diagonal with bi-residue $=1$ and no other pole.
In other words, this is the unique $1\boxtimes 1$ bidifferential on the annulus-square, that has a double pole on the diagonal with bi-residue $=1$ and no other pole, and that can be analytically continued to $\CC P^1\times \CC P^1\setminus \text{diagonal}$.

We normalize primitives at a fixed reference value $\xo$, independent of the moduli:
\begin{align}
L_0(x,\xi)&=\log\left(1-\frac{x-\xo}{\xi-\xo}\right),\\
L_+(x,\xi)&=L_0(x,\xi),\\
L_-(x,\xi)&=L_0(x,\xi)+R(x,\xi)-R(\xo,\xi).
\label{eq:normalized-L}
\end{align}
Thus $L_\pm(\xo,\xi)=0$ and
\begin{equation}
L_{+,x}(x,\xi)=\frac1{x-\xi},\qquad
L_{-,x}(x,\xi)=\frac1{x-\xi}+R_x(x,\xi).
\end{equation}
We then define the integration operators
\begin{align}
\mathcal L_\pm.\omega(z)&=-\frac1{2\pi\ii}
\oint_{\mathcal C_\pm}L_\pm(\x(z),\x(\zeta))\omega(\zeta). \cr 
\mathcal R_\pm.\omega(z)&= \frac1{2\pi\ii}
\oint_{\mathcal C_\pm} R(\x(z),\x(\zeta))\omega(\zeta). 
%\mathcal R_-.\omega(x)&=-\frac1{2\pi\ii}
%\oint_{\mathcal C_-}\bigl(R(x,\x(\zeta))-R(\xo,\x(\zeta))\bigr)\omega(\zeta),\cr
%\mathcal R_+.\omega(x)&=0.
\end{align}

\subsection{Resolvent}

We define $w_1(z) = W_{0,1}(\x(z))d\x(z)$, as a 1-form on $\curve_+\cup \mathcal D_+$.
It satisfies:
\bea
w_1(z) & \sim  & t \frac{dz}{z} \quad \text{ at } z\to\infty 
\eea
The discontinuity equation \eqref{eq:RMsdptequ} is then, on the cut  $z\in \mathcal C_0$:
\bea
dV(\x(z)) &=& w_1(z) + w_1(1/z) + \frac{d\x(z)}{2\pi\ii} \oint_{\mathcal C_+} R_x(\x(z),\x(z')) w_1(z') \cr
dV(\x(z)) &=& w_1(z) + w_1(1/z) + \sum_{j\in J} \alpha_j  w_1(\td\sigma_j(z)) \quad \text{under log hypothesis}.
\eea
This equation allows to analytically continue $w_1(z)$ to the whole annulus.
Therefore
\bea
w_1(z) & & \text{ is holomorphic in } 1/r<|z|<\infty  
\eea
and on the whole annulus it satisfies
\bea
dV &=& w_1+\mathfrak i^*w_1 + d\mathcal R_+.w_1 \cr 
dV &=& w_1+\mathfrak i^*w_1  + \sum_{j\in J} \alpha_j  \td\sigma_j^* w_1  \quad \text{under log hypothesis}.
\eea

\subsubsection{Rewriting with Cauchy kernels}

\bl
For $z\in\curve_+$, we have
\bea
&&  \frac{d\x(z)}{2\pi\ii} \oint_{\mathcal C_+} R_{x}(\x(z),\x(z')) \  w_1(z') \cr
&=&  \frac{d\x(z)}{2\pi\ii} \oint_{\mathcal C_+} \left( \frac{1}{\x(z)-\x(z')} + R_{x}(\x(z),\x(z')) \right)   w_1(z') \cr 
&=& \sum_{j\in J} \alpha_j  \td\sigma_j^*w_1(z)  \quad \text{under log hypothesis},
\eea
and
\bea
w_1(z)+w_1(1/z) 
&=& - \frac{d\x(z)}{2\pi\ii} \oint_{\mathcal C_+ + \mathcal C_-} \left( \frac{1}{\x(z)-\x(z')}+R_x(\x(z),\x(z')) \right) w_1(z')  .
\eea
%i.e.
%\bea
%w_1+\mathfrak i^* w_1  &=& d\mathcal L_+.w_1 + d\mathcal L_-.w_1 \ .
%\eea
We have 
\bea
\label{eq:VLcal}
%dV  &=& d\mathcal L_-.w_1 \cr 
%0 &=& d\mathcal L_+.w_1 \cr
V  &=& V(\xo) + \mathcal L_-.w_1 \cr 
0 &=& \mathcal L_+.w_1 .
\eea

\el

\proof{}
From our 1-cut assumption, for $z$ in the upper annulus, $w_1(z')/(\x(z)-\x(z'))$ is holomorphic in $\mathcal D_+\setminus \curve_+$, so its contour integral around $\mathcal C_+$ vanishes.
The integral on $\mathcal C_++\mathcal C_-=\partial\curve$ surrounds only the poles at $z'=z$ and $z'=1/z$ thus implying the second equality.
Adding the two gives
\bea
dV &=& d\mathcal L_-.w_1
\eea
and
\bea
0 &=& d\mathcal L_+.w_1.
\eea
Since $\mathcal L_\pm.w_1$ vanishes at $x=\xo$ by integration we get \eqref{eq:VLcal}.
\eproof

As an immediate corollary we have
\bp
Defining
\bea
V_+ &=& \mathcal L_+.w_1 \cr
V_- &=& \mathcal L_-.w_1 ,
\eea
we have
\bea
V_+ &=& 0 
\eea
and
\bea
V = V_- - V_+ +V(\xo).
\eea
\ep

\subsubsection{2 point function}

Define $w_2(z_1,z_2)$ such that in $(\curve_+\cup \mathcal D_+)\times (\curve_+\cup \mathcal D_+)$:
\bea
w_2(z_1,z_2) &=& W_{0,2}(\x(z_1),\x(z_2)) d\x(z_1) \otimes d\x(z_2).
\eea
It is symmetric, and bi-holomorphic in $(\curve_+\cup \mathcal D_+)\times (\curve_+\cup \mathcal D_+)$.

The "saddle-point" equation in random matrix theory, implies the following equation across the cut, i.e. for $z_1\in\mathcal C_0$:
\bea
 w_2(z_1,z_2)+w_2(1/z_1,z_2) && \cr 
 + \frac{d\x(z_1)}{2\pi\ii} \oint_{\mathcal C_+} R_x(\x(z_1),\x(\zeta)) w_2(\zeta,z_2) 
 &=& -  \frac{d\x(z_1)d\x(z_2)}{(\x(z_1)-\x(z_2))^2} 
 \eea
and the  symmetric equation for $z_2$.
Or under log hypothesis
 \bea
 w_2(z_1,z_2)+w_2(1/z_1,z_2) + \sum_j \alpha_j w_2(\td\sigma_j(z_1),z_2) 
 &=& -  \frac{d\x(z_1)d\x(z_2)}{(\x(z_1)-\x(z_2))^2} .
\eea
This equation allows to analytically continue $w_2$ to $\curve\times\curve$.
Because of the double pole in the right hand side, $w_2$ is meromorphic in $\curve\times\curve$, with a double pole at $z_1=1/z_2$, and no other pole.

We define
\bea
B(z_1,z_2) = w_2(z_1,z_2) +  \frac{d\x(z_1)\otimes d\x(z_2)}{(\x(z_1)-\x(z_2))^2}.
\eea
$B(z_1,z_2)$ is symmetric, and meromorphic in $\curve\times\curve$, with a double pole on the diagonal, with bi-residue=1.
And it can be analytically continued to $(\curve\cup \mathcal D_+)\times (\curve\cup \mathcal D_+)$ with also a double pole on the diagonal.

\bl
For any $z,z''\in \curve_+$, we have
\bea
0 &=& \frac{d\x(z)}{2\pi\ii}\oint_{\mathcal C_+} L_{+,x}(\x(z),\x(z')) B(z',z'') = - d\mathcal L_{+} .B  \cr
0 &=& \frac{d\x(z)}{2\pi\ii}\oint_{\mathcal C_-} L_{-,x}(\x(z),\x(z')) B(z',z'') = - d\mathcal L_{-} .B .
\eea
where $\mathcal L_\pm$ act on the first variable of $B$.
And we have
\bea
\mathcal L_{+} .B =0 \cr 
\mathcal L_{-} .B = 0 .
\eea
\el

\proof{}
$L_{+,x}(\x(z),\x(z')) w_2(z',z'') $ is holomorphic in $z'\in \mathcal D_+\setminus \curve_+$, therefore
\bea
\frac{1}{2\pi\ii}\oint_{\mathcal C_+} L_{+,x}(\x(z),\x(z')) w_2(z',z'')  
&=& 0.
\eea
The contour $\mathcal C_++\mathcal C_-$ surrounds the double pole of $w_2(z',z'')$ at $z'=1/z''$, and the poles of $L_{\pm,x}(\x(z),\x(z')$ at $z'=z,1/z$, which implies that
\bea
\frac{-d\x(z)}{2\pi\ii}\oint_{\mathcal C_++\mathcal C_-} L_{\pm,x}(\x(z),\x(z')) w_2(z',z'')  
&=& w_2(z,z'')+w_2(1/z,z'')  \cr 
&&  L_{\pm,x,x''}(\x(z),\x(z'')) d\x(z)d\x(z'') . \cr
\eea
Therefore, using the saddle-point equation we get
\bea
  -  \frac{d\x(z)d\x(z'')}{(\x(z)-\x(z''))^2} 
 &=& \frac{-d\x(z)}{2\pi\ii}\oint_{\mathcal C_-} L_{-,x}(\x(z),\x(z')) w_2(z',z'')  \cr 
&& - L_{-,x,x''}(\x(z),\x(z'')) d\x(z)d\x(z'') 
\eea
In the right hand side, changing from $w_2$ to $B$ gives
\bea
  -  \frac{d\x(z)d\x(z'')}{(\x(z)-\x(z''))^2}
 &=& \frac{-d\x(z)}{2\pi\ii}\oint_{\mathcal C_-} L_{-,x}(\x(z),\x(z')) B(z',z'')  \cr 
 && - \frac{d\x(z)}{2\pi\ii}\oint_{\mathcal C_-} L_{-,x}(\x(z),\x(z'))  \frac{d\x(z')d\x(z'')}{(\x(z')-\x(z''))^2} \cr 
&& - L_{-,x,x''}(\x(z),\x(z'')) d\x(z)d\x(z'')  \cr 
 &=& \frac{-d\x(z)}{2\pi\ii}\oint_{\mathcal C_-} L_{-,x}(\x(z),\x(z')) B(z',z'')  \cr 
 && -  \frac{d\x(z)d\x(z'')}{(\x(z)-\x(z''))^2} \cr 
\eea
Therefore the saddle point equation for $w_2$ implies
\bea
0 &=& \frac{d\x(z)}{2\pi\ii}\oint_{\mathcal C_-} L_{-,x}(\x(z),\x(z')) B(z',z'') ,
\eea
and we already have
\bea
0 &=& \frac{d\x(z)}{2\pi\ii}\oint_{\mathcal C_+} L_{+,x}(\x(z),\x(z')) B(z',z'')  .
\eea
By integrating and since the result vanishes at $\zo$ we get
\bea
\mathcal L_{+} .B =0 \cr 
\mathcal L_{-} .B = 0.
\eea
\eproof

\subsubsection{Formulation as generalized cycles}

For $x\in \x(\curve)$, the map which to a meromorphic differential $\omega$ associates $2\pi\ii\mathcal L_\pm(x).\omega$
\bea
\acycle_{\pm}(x)  = 2\pi\ii \mathcal L_{\pm}(x)
\eea
belongs to the dual of the space of meromorphic forms on $\curve$:
\bea
\acycle_{\pm}(x) \in \mathfrak M^1(\curve)^*.
\eea
Moreover, its pairing to $B$ vanishes, and since $0$ belongs to the vector space $\mathfrak M^1(\curve)$, this implies that for each $x\in\x(\curve)$, $\acycle_{\pm}(x)$ is a generalized cycle in the sense of \cite{EynGenCycles}:
\beq
\acycle_{\pm}(x) \in \Ker \hat B \subset \mathfrak M_1(\curve) \subset \mathfrak M^1(\curve)^*
\eeq
where $\hat B$ is the map from generalized cycles to meromorphic 1-forms, by pairing with $B$. 
Here we have $\hat B(\acycle_{\pm}(x))=0$, therefore $\acycle_{\pm}(x)\in \Ker\hat B$.
The pairing between generalized cycles and meromorphic forms is usually denoted, as the Poincarr\'e pairing, as an integral:
\beq
<\gamma,\omega> \overset{{\text{notation}}}{=} \oint_\gamma \omega.
\eeq

We thus have
\begin{equation}
\frac1{2\pi\ii}\oint_{\acycle_\pm(x)}w_1=V_\pm(x),\qquad
\oint_{\acycle_\pm(x)}B=0
\end{equation}
and the matrix model potential is $V=V_- - V_+ + V(\xo)$.

%
%One can write the Taylor expansion of $\acycle_\pm(x)$ near $x=\xo$, and get
%\bea
%\acycle_\pm(x) = \sum_{k=1}^\infty \frac{1}{k} (x-\xo)^k \acycle_{\pm,k}.
%\eea
%This gives:
%\begin{align}
%\acycle_{+,k}&=\mathcal C_+\cdot(\xi-\xo)^{-k},\\
%\acycle_{-,k}&=\mathcal C_-\cdot\left((\xi-\xo)^{-k}
%-\frac{\partial_x^kR(\xo,\xi)}{(k-1)!}\right).
%\end{align}
%In the  log-hypothesis this becomes
%\begin{equation}
%\acycle_{-,k}=\mathcal C_-\cdot\left((\xi-\xo)^{-k}
%+\sum_j\alpha_j(\sigma_j(\xi)-\xo)^{-k}\right).
%\label{eq:root-cycles}
%\end{equation}
%

Let us assume here that $R(x,x')$ can be decomposed on the basis of Chebyshev polynomials ($T_k((z+1/z)/2) = (z^k+z^{-k})/2$):
\bea
R(x,x') &=& \sum_{k,l\geq 1} \frac{R_{k,l}}{kl} \  \ 4  T_k((x-\alpha)/2\beta)T_l((x'-\alpha)/2\beta) \cr
 &=& \sum_{k,l\geq 1} \frac{R_{k,l}}{kl} \  \  (z^k+z^{-k})(z'^l+z'^{-l})
\eea
(where $x=\x(z)$ and $x'=\x(z')$),
where $R=R^T$ is a symmetric infinite matrix.
We have:

\bl[Expansion on the Chebyshev basis]
\label{Lem:Tcehbychevbasis}
If $z\in\curve$ we have
\bea
\acycle_{-}(x)
&=& - \sum_{k\geq 1} (2T_k((x-\alpha)/2\beta)-2T_k((\xo-\alpha)/2\beta)) \ \acycle_{-,k} \cr 
 \acycle_{-,k} &=& {\mathcal C_-}. \left( \frac{-1}{k} \zeta^k +  \sum_{l\geq 1}  \frac{1}{kl} R_{k,l}  (\zeta^l+\zeta^{-l}) \right).
\eea

\el
\proof{}
Observe that $(\x(z)-\x(\zeta)) = \beta (z-\zeta)(1-1/z\zeta)$.

In the integration we have $\zeta\in\mathcal C_-$, and since $z\in \curve$, therefore  $|\zeta|< \min(|z|,1/|z|)$ and we have the absolutely convergent expansion
\bea
\ln{(\x(z)-\x(\zeta))/(\xo-\x(\zeta))}
&=& - \sum_{k=1}^\infty \frac{1}{k} \zeta^k \ \left(z^k+z^{-k} -z_0^k - z_0^{-k}  \right) .
\eea
The coefficient of $2T_k((x-\alpha)/2\beta)-2T_k((\xo-\alpha)/2\beta)$ in $L_-(x,\x(\zeta))$ is thus:
\bea
{\mathcal C_-}. \left( \frac{-1}{k} \zeta^k +  \sum_{l\geq 1}  \frac{1}{kl} R_{k,l}  (\zeta^l+\zeta^{-l}) \right).
\eea
\eproof

In the case $R=0$, remark that:
\bp[Trivial interaction $R=0$ case]
If $R=0$, a basis of cycles annihilating $B$ is given by the same cycles as in the GUE case:
\bea
\td\acycle_{+,k} &=& \mathcal C_+. (\x-\xo)^{-k} \cr 
\td \acycle_{-,k} &=& \mathcal C_-.  (\x-\xo)^{-k} .
\eea
It is easy to verify that the fundamental 2nd kind form of $\CC P^1$,   $\td B(z_1,z_2) = \frac{dz_1 \otimes dz_2}{(z_1-z_2)^2}$, satisfies 
\bea
\oint_{\td\acycle_{\pm,k}} \td B = 0
\eea
i.e. $B=\td B$ is solution of the saddle point equations.
In other words, when $R=0$, we recover the well known result that $B=\td B$.
\ep
\proof{Easy computation.}\eproof

\section{From spectral curve to matrix model}

%================================================
%================================================

%\reviewnote{This section remains a working reconstruction argument. The residue identities for $S$ below are established, but the general converse, descent to a symmetric interaction, analytic invertibility, the construction of the physical resolvent, and the higher-genus identification still need proof. The proposed formulas explicitly marked below are not asserted as final results.}
%
%The stated local data need an additional physical admissibility condition. For example, with fixed Joukowsky $x$, the bidifferential
%\begin{equation}
%B(z,z')=\td B(z,z')+c\,dz\,dz',\qquad c\ne0,
%\end{equation}
%is symmetric, has the required diagonal pole and zero boundary periods. Nevertheless its connected scalar covariance does not decay at the physical infinity. Pairing it with a Gaussian one-form of positive period gives data satisfying the local conditions below, but not the planar covariance of a matrix model with that same fixed $x$. A change of polarization would have to be specified and justified; it cannot be implicit.
%

The goal here is to do the converse: given a spectral curve with 2 branch points in a cylinder, is there a matrix model with a measure of the form \eqref{eq:defRMmeasure} that has that spectral curve.
In other words, recover the potential $V(x)$ and the 2-particle interaction $R(x,x')$ from the spectral curve.

\subsection{Spectral curve}

Consider a spectral curve of the following form
\bea
\spcurve=
\begin{cases}
\curve = \CC \cap \{1/r<|z|<r\} \quad = \ \text{cylinder} \cr 
\x(z) = \alpha + \beta (z+1/z) \cr 
\y(z) = \text{holomorphic 1-form on }\curve \ \ \text{such that } \y(1)=\y(-1)=0 \cr 
\quad \quad = y(z)d\x(z) \cr
B(z_1,z_2) = \frac{1}{(z_1-z_2)^2} dz_1 \otimes dz_2 + \text{holomorphic on }\curve\times \curve \cr 
\qquad \text{such that } \ \ B(z_2,z_1)=B(z_1,z_2) \cr 
\qquad \text{and } \ \int_{\mathcal C_\pm} B(z,.)=0
\end{cases}
\eea
We define the same geometry as in section \ref{sec:geometryCylinder}.

Moreover, up to shrinking the radius of the annulus, we assume that $\y$ and $B$ can be analytically continued to a 2-sided neighborhood of the boundary $\partial \curve$, so that integrals on $\mathcal C_\pm$ will be well defined.

It is important for what follows that $B$ has no holonomy around the cylinder:
\beq
\forall z\in \curve \ , \quad \int_{\mathcal C_\pm} B(z,.)=0.
\eeq

Then we define:
\bea
\label{eq:deft}
t = \frac{1}{2\pi\ii} \oint_{\mathcal C_+} \y = -\frac{1}{2\pi\ii} \oint_{\mathcal C_-} \y .
\eea 
We define also $F_{0,1}$ such that 
\beq
dF_{0,1}=\y
\eeq
If $t\neq 0$, $\y$ has some holonomy around the cylinder, and  $F_{0,1}$ can't be analytic on the whole cylinder.
We define it with a cut on $\RR_-\cap\curve$.
The choice of integration constant here is arbitrary. We shall normalize it so that 
\beq
F_{0,1}(\zo)=0.
\eeq

\subsection{GUE case}

It is useful to denote
\bd[GUE kernels]
\bea
\td B(z,z') &=& \frac{dz \otimes dz'}{(z-z')^2} \cr 
\td S(z;z') &=& \left(\frac{1}{z-z'}-\frac{1}{z-\zo} \right)dz \cr 
\td E(z,z') &=& \frac{z-z'}{\sqrt{dz dz'}}.
\eea 
so that we have
\bea
\td S(z;z') &=& d_z \ln\frac{\td E(z,z')}{\td E(z,\zo)} \cr 
\td B(z,z') &=& d_{z'} \td S(z;z') = d_z d_{z'} \ln{\td E(z,z')} .
\eea
\ed

\bd[GUE generalized cycles]
We define the generalized 2nd kind cycles (boundary cycles in \cite{EynGenCycles}) as:

$\bullet$ for $k\geq 0$:
\bea
\td\acycle_{\pm,k} &=& \mathcal C_\pm. (\x-\xo)^{-k} ,
\eea
$\bullet$ for $k> 0$:
\bea
\td\bcycle_{\pm,k} &=& \frac{1}{2\pi\ii k} \mathcal C_\pm. (\x-\xo)^{k} 
\eea
$\bullet$ and for $k=0$:
\bea
\td\bcycle_{+,0} &=& \gamma_{\zo\to -r} + \frac{1}{2\pi\ii} \gamma_{z=r e^{\ii\theta}, \ \theta=-\pi\to\pi} . \ln{(\x-\xo)} \cr 
\td\bcycle_{-,0} &=& \gamma_{\zo\to -1/r} - \frac{1}{2\pi\ii} \gamma_{z=r^{-1} e^{\ii\theta}, \ \theta=-\pi\to\pi} . \ln{(\x-\xo)}.
\eea

\ed

%TODO move that elsewhere ?
%
%For example, for the GUE case with $\x(z)=z+1/z$, $y(z)=t/z$ and $\xo=0$ we would have
%\bea
%\int_{\td\bcycle_{+,0}} \y 
%&=& \int_{\ii}^{-r} \y + \frac{1}{2\pi\ii} \int_{z=r e^{\ii\theta}, \ \theta=-\pi}^\pi  \y(z) \ln{(\x(z)-\xo)}  \cr 
%&=& \int_{\ii}^{-r} \y + \frac{1}{2\pi\ii} \int_{z=r e^{\ii\theta}, \ \theta=-\pi}^\pi  (\y(z)-t d\x(z)/\x(z)) \ln{(\x(z)-\xo)}  \cr 
%&& +   \frac{t}{2\pi\ii} \int_{z=r e^{\ii\theta}, \ \theta=-\pi}^\pi \frac{d\x(z)}{\x(z)} \ln{(\x(z)-\xo)} \cr 
%\eea

\subsection{Kernels}

\bd
Define, for $\epsilon=\pm$ and $\eta=\pm$:
\bea
B^{\epsilon,\eta} = \td B + (1-\Pi^\epsilon)\otimes (1-\Pi^\eta).(B-\td B) .
\eea
From lemma \ref{Lem:Pipmextend},  $B^{\epsilon,\eta}(z_1,z_2)$
can be analytically extended to a meromorphic bidifferential on $(\curve\cup\mathcal D_\epsilon)\times (\curve\cup\mathcal D_\eta)$ having a double pole on the diagonal and no other pole.
And in $\curve\times \curve$ it has the same odd part as $B$:
\bea
(1-\mathfrak i^*)\otimes (1-\mathfrak i^*) B^{\epsilon,\eta} &=& (1-\mathfrak i^*)\otimes (1-\mathfrak i^*)  B  .
\eea
\ed

\bd
Define
\bea
\check B(z_1,z_2) = B^{+,+}(z_1,z_2) \cr
w_2(z_1,z_2) = \check B(z_1,z_2) - \frac{d\x(z_1)\otimes d\x(z_2)}{(\x(z_1)-\x(z_2))^2}.
\eea
\ed

\bp
$\check B$ and $w_2$ are symmetric and differ from $B$ by  terms even in at least one of the 2 variables, 
$w_2$  is bi-holomorphic in $(\curve_+\cup\mathcal D_+)\times (\curve_+\cup\mathcal D_+)$, and $\check B$ is meromorphic in the same domain with a normalized double pole on the diagonal and no other pole.
Moreover they have no monodromy around the cylinder.
\ep
\proof{Easy.}\eproof

\bl
Let $\dot B$ be one of $B, B^{+,+},B^{-,-}$
For $z,z'\in \curve$, we have
\beq
\dot  B(z,z') = \td B(z,z') 
+ \frac{1}{(2\pi\ii)^2}\sum_{\epsilon=\pm 1} \oint_{z_2\in\mathcal C_\epsilon} \oint_{z_1\in\mathcal C_\epsilon} \frac{dz}{z-z_1} \frac{d\x(z_1)}{\x(z_1)-\x(z_2)} \dot  B(z_2,z')
\eeq
where in the integration we take $|z_2^\epsilon|>|z_1^\epsilon|$.
As a corollary we have
\bea
\frac{d\x(z) d\x(z')}{(\x(z)-\x(z'))^2}
&=& \dot  B(z,z')+\dot  B(1/z,z') \cr 
&& - \frac{1}{(2\pi\ii)^2}\sum_{\epsilon=\pm 1} \oint_{z_2\in\mathcal C_\epsilon} \oint_{z_1\in\mathcal C_\epsilon} \frac{d\x(z)}{\x(z)-\x(z_1)} \frac{d\x(z_1)}{\x(z_1)-\x(z_2)} \dot  B(z_2,z') \cr
&=& \dot  B(z,z')+\dot  B(1/z,z') \cr 
&& + \frac{1}{2\pi\ii} \oint_{z_2\in\mathcal C}  \frac{d\x(z)}{\x(z)-\x(z_2)} \dot  B(z_2,z') .
\eea

\el
\proof{}
Same proof as in \cite{EOBKMP}.
We have
\bea
&& B(z,z') - \td B(z,z') \cr 
&=& \Res_{z_1\to z} \frac{dz}{z_1-z} \left( B(z_1,z') - \td B(z_1,z') \right) \cr 
&=& \frac{1}{2\pi\ii}\sum_{\epsilon=\pm 1} \oint_{z_1\in\mathcal C_\epsilon} \frac{dz}{z_1-z} \left( B(z_1,z') - \td B(z_1,z') \right) \cr 
&=& \frac{1}{2\pi\ii}\sum_{\epsilon=\pm 1} \oint_{z_1\in\mathcal C_\epsilon} \frac{dz}{z_1-z} \Res_{z_2\to {z_1}} \frac{d\x(z_1)}{\x(z_2)-\x(z_1)} \left( B(z_2,z') - \td B(z_2,z') \right) \cr 
&=& \frac{1}{(2\pi\ii)^2}\sum_{\epsilon=\pm 1} \oint_{z_2\in\mathcal C_\epsilon} \oint_{z_1\in\mathcal C_\epsilon, \ |z_1^\epsilon|<|z_2^\epsilon|} \frac{dz}{z_1-z}  \frac{d\x(z_1)}{\x(z_2)-\x(z_1)} \left( B(z_2,z') - \td B(z_2,z') \right) \cr 
&& - \frac{1}{(2\pi\ii)^2}\sum_{\epsilon=\pm 1} \oint_{z_2\in\mathcal C_\epsilon} \oint_{z_1\in\mathcal C_\epsilon, \ |z_1^\epsilon|>|z_2^\epsilon|} \frac{dz}{z_1-z}  \frac{d\x(z_1)}{\x(z_2)-\x(z_1)} \left( B(z_2,z') - \td B(z_2,z') \right) \cr 
\eea
In the last line, we can move the integration contour of $z_1$ to $z_1^\epsilon\to \infty$, and since this encloses at most the simple pole at $\infty$, whose residue gives a constant and since the period of $B-\td B$ vanishes,  we get
\bea
&& B(z,z') - \td B(z,z') \cr 
&=& \frac{1}{(2\pi\ii)^2}\sum_{\epsilon=\pm 1} \oint_{z_2\in\mathcal C_\epsilon} \oint_{z_1\in\mathcal C_\epsilon, \ |z_1^\epsilon|<|z_2^\epsilon|} \frac{dz}{z_1-z}  \frac{d\x(z_1)}{\x(z_2)-\x(z_1)} \left( B(z_2,z') - \td B(z_2,z') \right) \cr 
\eea
For the term with $\td B(z_2,z')$ we can also move the integration contour of $z_2$ to $z_2^\epsilon\to \infty$ without encircling a pole, so the the integral vanishes. 
This proves the lemma.
\eproof{}

\subsubsection{Third-kind differential}

\bd
For any $\dot B\in \{B,B^{+,+},B^{-,-}\}$, since $\dot B$ has no holonomy around the cylinder, we can define a 1-form $\dot S(z;z')$ whose differential in the 2nd variable is $\dot B(z,z')$:
\bea
\dot S(z;z') = \int_{z''=\zo}^{z'} \dot B(z,z'').
\eea 
It satisfies:
\bea
d_{z'} \ \dot S(z;z') &=& \dot B(z,z') \cr 
\dot S(z;\zo) &=& 0 \cr 
\Res_{z\to z'} \dot S(z;z') &=& 1 \cr 
\Res_{z\to \zo} \dot S(z;z') &=& -1 .
\eea
$\dot S(z;z')$ is a meromorphic 1-form in the variable $z$, and a meromorphic function of $z'$, it has a simple pole at $z=z'$ with residue 1 and a simple pole at $z=\zo$ with residue $-1$ (except when $z'=\zo$ at which the poles cancel).
\bea
\dot S\in H^0(\curve\times\curve , K_\curve(\diag{} + \zo))_{\Res=+1,-1} .
\eea 
In other words
\bea
\dot S(z;z') - \left(\frac{1}{z-z'}-\frac{1}{z-\zo} \right)dz = \text{ holomorphic on } \curve\times \curve.
\eea
\ed

Moreover:
$S^{+,+}$ (resp. $S^{-,-}$) can be analytically extended to $(\curve\cup \mathcal D_+)\times (\curve\cup \mathcal D_+)$ (resp. $(\curve\cup \mathcal D_-)\times (\curve\cup \mathcal D_-)$) to have the same pole properties.

\subsection{Generalized cycles}

\bl[Generating function of $\acycle$-cycles]
Let
\bea
\acycle(z) &=& -(\mathcal C_++\mathcal C_-).\check S(z;.) .
\eea 
For any $z,z''\in\curve$ it satisfies
\bea
\frac{1}{2\pi\ii}\oint_{\acycle(z)} \check B(.,z'') &=& 0.
\eea

\el
\proof{}
This lemma is in fact a direct application of \cite{EynGenCycles}, but let us reprove it here in this specific setting:
\bea
\frac{1}{2\pi\ii}\oint_{\acycle(z)} \check B(.,z'')
&=& \frac{-1}{2\pi\ii} \oint_{z'\in \mathcal C_+ +\mathcal C_-} S(z;z')\check B(z',z'') \cr
&=& -\Res_{z'=z} S(z;z')\check B(z',z'') - \Res_{z'=z''} S(z;z')\check B(z',z'') \cr 
&=& \check B(z,z'') - \check B(z,z'') \cr 
&=& 0.
\eea
\eproof

\bl
If $\omega$ is a holomorphic 1-form on $\curve$ we have
\bea
\frac{1}{2\pi\ii}\oint_{\acycle(z)} \omega &=& \omega(z)
\eea
\el
\proof{}
This is also a direct application of \cite{EynGenCycles}, and it is in fact the Riemann bilinear identity. 
Let us reprove it here in this specific setting:
\bea
\frac{1}{2\pi\ii}\oint_{\acycle(z)} \omega 
&=& \frac{-1}{2\pi\ii} \oint_{z'\in \mathcal C_+ +\mathcal C_-} S(z;z')\omega(z') \cr
&=& -\Res_{z'=z} S(z;z')\omega(z') \cr
&=& \omega(z)  .
\eea
\eproof

\bl[Trivial upper side cycles]
For the kernel $\check B$ we have
for $k\geq 0$
\bea
\oint_{\td{\mathcal A}_{+,k}} \check B=0.
\eea
In other words the GUE upper side cycles also annihilate $\check B$:
\bea
\forall k\geq 0 , \qquad \acycle_{+,k} = \td{\acycle}_{+,k}  \qquad \text{annihilates } \check B.
\eea
\el
\proof{Indeed, $\check B(z,z'')$ can be analytically continued to $z''\to \infty$, $(\x(z'')-\xo)^{-k}$ is also holomorphic in $\mathcal D_+\setminus\curve_+$, therefore the integral of $\td{\mathcal A}_{+,k}$ encloses no pole and thus vanishes.\eproof
}

It remains to find the cycles that annihilate $\check B$ on the lower side of the cylinder, in general they differ from those of GUE.

We decompose $\check B(z,\zeta)$ with lemma \ref{Lem:Laurentdecomp}:
\bea
\check B(z,\zeta) - \td B(z,\zeta)
&=& \sum_{k,l\geq 1} \frac{1}{kl} H_{k,l}  \ dz^{-k} \otimes d\zeta^{-l} .
\eea
which implies
\bea
\check S(z,\zeta) - \td S(z,\zeta)
&=& \sum_{k,l\geq 1}  \frac{1}{kl} H_{k,l}  (\zeta^{-l}-\zo^{-l})  dz^{-k}  .
\eea
Since $\check B-\td B$ is analytic in the cylinder, this implies that there exists some $C>0$ such that
\beq
\label{eq:boundHkl}
\forall k,l \ , \qquad |H_{k,l}| < C \ r^{-k-l} .
\eeq

For $|\zeta|<|z|$ this gives
\bea
\check B(z,\zeta) 
&=& \sum_{k\geq 1} \frac{-1}{k} d(z^{-k})  \otimes \left(  d\zeta^k - \sum_{l\geq 1} \frac{1}{l}  H_{k,l}  d\zeta^{-l}\right) \cr
\check S(z,\zeta) 
&=& \sum_{k\geq 1} \frac{-1}{k} d(z^{-k})  \left( (\zeta^k-\zo^k) - \sum_{l\geq 1}  \frac{1}{l} H_{k,l}  (\zeta^{-l}-\zo^{-l}) \right) \cr
&=& \sum_{k\geq 1} \frac{-1}{k} d(z^{-k})  \left( \zeta^k - \sum_{l\geq 1}  \frac{1}{l} H_{k,l}  \zeta^{-l} \right)  + \text{independent of }\zeta.
\eea
Therefore we have:
\bea
\forall k\geq 1 \qquad
\oint_{\mathcal C_-}  \left( \frac{-1}{k} \zeta^k +  \sum_{l\geq 1}  \frac{1}{kl} H_{k,l}  \zeta^{-l} \right) \check B(\zeta,z') = 0.
\eea

This implies
\bp[Lower side cycles]
The lower side cycles that annihilate $\check B$ are generated by the family
\bea
\label{eq:AcycleschekB}
k\geq 1 \qquad
\mathcal C_- .  \left( \frac{-1}{k} \zeta^k +  \sum_{l\geq 1}  \frac{1}{kl} H_{k,l}  \zeta^{-l} \right) .
\eea
\ep
Notice that the cycle $\mathcal C_-$ also annihilates $\check B$, but since it is isotopic to $-\mathcal C_+$, we may consider that it is already in the family of upper cycles.

\subsection{Recovering the 2-body potential $R(x,x')$}

Now, let us look for an annihilating cycles of the form of lemma~\ref{Lem:Tcehbychevbasis}, i.e.
\bea
\label{eq:AcycleskAnstazR}
{\mathcal C_-}. \left( \frac{-1}{k} \zeta^k +  \sum_{l\geq 1}  \frac{1}{kl} R_{k,l}  (\zeta^l+\zeta^{-l}) \right).
\eea
The goal is to find the symmetric matrix $R_{k,l}$ so as to match \eqref{eq:AcycleschekB}.

\bt[2-body interaction]
If a solution exists, the matrix $R$ is given by
\bea
\label{eq:solR}
R  &=&  H  (1+K^{-1}H )^{-1} \cr 
  &=&  H  - HK^{-1}H + HK^{-1}HK^{-1}H - \dots 
\eea
The 2-body interaction is then
\bea
R(x,x') &=& \sum_{k,l} \frac{R_{k,l}}{kl} \  \  4 T_k((x-\alpha)/2\beta)T_l((x'-\alpha)/2\beta) .
\eea
A sufficient condition for a solution of \eqref{eq:AcycleskAnstazR}   to exist is 
\bea
\label{eq:boundCcondition}
C < \frac{1}{|\ln{(1-r^{-2})}|} \qquad
\text{or} \qquad
r> \frac{1}{\sqrt{1-e^{-1/C}}}.
\eea
When this condition is satisfied the series \eqref{eq:solR} is absolutely convergent.
\et

\proof{} 
Writing the vectors $u_+=(\zeta,\zeta^2,\zeta^3,\dots)$ and $u_-=(\zeta^{-1},\zeta^{-2},\zeta^{-3},\dots)$, and $K=\diag(1,2,3,\dots)$, we see that the vector space of annihilators of $\check B$ is generated by the vectors
\beq
u_+ - H K^{-1} u_-
\eeq
while the vector space of annihilators of the form \eqref{eq:AcycleskAnstazR} is generated by the vectors
\beq
u_+ -  R K^{-1} (u_++u_-)
\eeq
Observe that
\beq
(1-R K^{-1})(u_+ -  H K^{-1} u_-) = (1-R K^{-1} )u_+ - (1-R K^{-1})  H K^{-1} u_-.
\eeq
Therefore we need to have
\bea
R K^{-1} &=& (1-R K^{-1}) H K^{-1} \cr 
K^{-\frac12} R K^{-\frac12} &=& (1-K^{-\frac12}R K^{-\frac12}) K^{-\frac12} H K^{-\frac12} \cr 
K^{-\frac12} R K^{-\frac12} (1+K^{-\frac12}H K^{-\frac12}) &=& K^{-\frac12} H K^{-\frac12} \cr 
K^{-\frac12} R K^{-\frac12}  &=& K^{-\frac12} H K^{-\frac12} (1+K^{-\frac12}H K^{-\frac12})^{-1} \cr 
R  &=&  H  (1+K^{-1}H )^{-1} \cr 
R  &=&  H  - HK^{-1}H + HK^{-1}HK^{-1}H - \dots \cr 
\eea
The bound \eqref{eq:boundHkl} implies 
\beq
|R_{k,l}| \leq C r^{-k-l} \ \frac{1}{1+C\ln{(1-r^{-2})}}
\eeq
Therefore, the series is absolutely convergent when
\bea
C < \frac{1}{|\ln{(1-r^{-2})}|} < r^2 \qquad
\text{or} \qquad
r> \frac{1}{\sqrt{1-e^{-1/C}}}.
\eea

Eventually this defines
\bea
R(x,x') &=& \sum_{k,l} \frac{R_{k,l}}{kl} \  \  4T_k((x-\alpha)/2\beta)T_l((x'-\alpha)/2\beta) 
\eea
The bound \eqref{eq:boundHkl} and \eqref{eq:boundCcondition} implies that the series is absolutely convergent for $x,x'\in \x(\curve)$.
\eproof{}

\textbf{Remark:} The condition \eqref{eq:boundCcondition} is a sufficient condition for the solution to exist, it is not a necessary condition, but it is important. When this bound is not satisfied, the solution may not exist. This can be seen on the following axample:
choose
\beq
B(z_1,z_2)  = \td B(z_1,z_2) + c \frac{dz_1}{z_1^{2}}\otimes \frac{dz_2}{z_2^{2}}.
\eeq
It has $H_{1,1}=c$ and all other $H_{k,l}=0$.
The bounding constant $C$ must  be chosen such that
\beq
C\geq |c| \ r^2
\eeq
and \eqref{eq:boundCcondition} is thus the condition that
\bea
|c| r^2 < \frac{1}{|\ln{(1-r^{-2})|}} < r^2
\eea
in particular it requires
\beq
|c|<1.
\eeq

Let us see what happens if this condition is not satisfied:
$H$ is a rank 1 matrix
\beq
H = c e_1 e_1^T
\eeq
where $e_1=(1,0,0,\dots)$.
The equation $R(1+K^{-1}H)=H$ gives
\bea
R(1+c e_1 e_1^T) = c e_1 e_1^T
\eea
The symmetric solution is:
\beq
R = \frac{c}{1+c} \  e_1 e_1^T.
\eeq
If $c=-1$ there is no solution.

\subsection{Recovering the 1-body potential $V(x)$}

\bd
we define
\beq 
\x^* dV_+ = \Pi^+.\y
\eeq
and we normalize the primitive $V_+$ so that $V_+(\xo)=0$.
We also define
\bea
w_1 = \y - \x^*dV_+ = \y - \Pi^+.\y.
\eea
Then, use $R(x,x')$ and the corresponding operator $\mathcal L_-(x)$ to define
\bea
V(x) = \mathcal L_-(x). w_1.
\eea

\ed

We have the following properties
\begin{itemize}
\item $w_1$ can be analytically continued to a 1-form in $\curve\cup \mathcal D_+$, that has a simple pole at $z=\infty$.
\item At $\infty$ 
\beq
w_1 \sim t \frac{d\x(z)}{\x(z)} (1+O(1/\x(z))).
\eeq
\item $\y$ and $w_1$ have the same odd part, they differ by an even part.
\end{itemize}

\subsection{Topological Recursion}

In this section, we assume that the spectral curve is regular, i.e. $y=\y/dx$ is finite at the branch points, and its derivative $y'(\pm 1)$ is not vanishing. With these hypothesis,
Topological Recursion \cite{EO07} assigns to the spectral curve, a sequence of differential forms $\omega_{g,n}$.

Consider the two spectral curves $\spcurve=(\curve,\x,\y,B)$ and $\check\spcurve=(\curve,\x,w_1,\check B)$.
Let us denote
\bea
\omega_{g,n} &=& \omega_{g,n}(\spcurve) \cr 
\check\omega_{g,n} &=& \omega_{g,n}(\check\spcurve) .
\eea

\bt
For $2g-2+n>0$, $\omega_{g,n}$ and $\check\omega_{g,n}$ have the same totally odd part:
\bea
(1-\mathfrak i^*)\otimes \dots (1-\mathfrak i^*) \check\omega_{g,n} = (1-\mathfrak i^*)\otimes \dots (1-\mathfrak i^*) \omega_{g,n}.
\eea

In particular the moments ($k\geq 0$) are the same:
\beq
\frac{1}{2\pi\ii} \oint_{\mathcal C_+} \omega_{g,1}(z)  \  \x(z)^{k} 
= \frac{1}{2\pi\ii} \oint_{\mathcal C_+} \check\omega_{g,1}(z)  \  \x(z)^{k}.
\eeq
and more generally for $k_i\geq 0$:
\bea
\frac{1}{(2\pi\ii)^n} \oint_{\mathcal C_+}\dots\oint_{\mathcal C_+}\  \omega_{g,n}(z_1,\dots,z_n)  \  \x(z_1)^{k_1}  \dots \x(z_n)^{k_n}  \cr 
= \frac{1}{(2\pi\ii)^n} \oint_{\mathcal C_+}\dots \oint_{\mathcal C_+} \ \check\omega_{g,n}(z_1,\dots,z_n)  \  \x(z_1)^{k_1}  \dots \x(z_n)^{k_n}.
\eea

\et

\proof{}
The Topological recursion  kernel is 
\bea
K(z,z') = \frac{\int_{\mathfrak i(z')}^{z'} B(z,.)}{2(\y(z)-\mathfrak i^*\y(z))}
\eea
it clearly involves only the odd part of $\y$, and it involves the odd part of $B$ in the second variable.
Taking also the odd part in the first variable implies that
\bea
(1-\mathfrak i^*) \check K = (1-\mathfrak i^*) K
\eea
and therefore the odd part in all variables of stable $\omega_{g,n}$ are the same for $\spcurve$ and $\check \spcurve$.

For the moments, TR implies (\cite{EO07}) that, for $g\geq 1$, $\omega_{g,1}+\mathfrak i^*\omega_{g,1}$ has no pole at the ramification points, therefore
\bea
\oint_{\mathcal C_++\mathcal C_-} (\omega_{g,1}+\mathfrak i^*\omega_{g,1}) \x^k = 0
\eea
Since $\mathfrak i.\mathcal C_- = \mathcal C_+$ this implies
\bea
\oint_{\mathcal C_+} (\omega_{g,1}+\mathfrak i^*\omega_{g,1}) \x^k = 0
\eea
and thus
\bea
\oint_{\mathcal C_+} \omega_{g,1} \x^k = \frac12  \oint_{\mathcal C_+} (\omega_{g,1}-\mathfrak i^*\omega_{g,1}) \x^k .
\eea
Since this depends only on the odd part, it is the same for $\omega_{g,1}$ and $\check\omega_{g,1}$.
This extends to all stable $\omega_{g,n}$.
\eproof

%================================================
%================================================

\section{Recap}

%================================================
%================================================

To identify with the matrix model we do the following steps:
\begin{itemize}
\item Define the kernel $\check B$:
\bea
\check B &=& \td B + (1-\Pi^+)\otimes (1-\Pi^+)(B-\td B) .
\eea

\item After subtracting the GUE double pole, expand it as
\bea
\check B(z,z') &=& \td B(z,z') +  \sum_{k,l\geq 1} \frac{H_{k,l}}{kl} dz^{-k} \otimes d z'^{-l}.
\eea

\item  define the matrix $R$ as 
\bea
R = H (1+K^{-1}H)^{-1} = H - HK^{-1} H + H K^{-1} H K^{-1} H + \dots
\eea
(where $K=\diag(1,2,3,...)$),

\item The 2-body interaction is then
\bea
R(x,x') &=&  \sum_{k,l\geq 1} \frac{R_{k,l}}{kl} \ (z^k+z^{-k})(z'^l+z'^{-l}) \cr 
&=&  \sum_{k,l\geq 1} \frac{R_{k,l}}{kl} \ 4T_k((x-\alpha)/2\beta) T_l((x'-\alpha)/2\beta) .
\eea

\item Then 1-body potenital is
\bea
V(x) &=& \mathcal L_-(x).w_1.
\eea

\item The matrix model 1-point and 2-point function are
\bea
w_1 &=& \y - \Pi^+\y \cr 
w_2 &=& \check B - \frac{dx_1\otimes dx_2}{(x_1-x_2)^2} .
\eea

\item Then, choosing $\hbar=t/N$ for some strictly positive integer $N$, we can define the formal random matrix measure
\beq
\label{eq:defRMmeasure2}
d\mu(\Lambda) = \frac{1}{N! Z_N(V,R)} \ 
d\Lambda_1\dots d\Lambda_N
\ \prod_{i<j} (\Lambda_i-\Lambda_j)^2 \prod_{i,j} e^{\frac12 R(\Lambda_i,\Lambda_j)} e^{- \hbar^{-1} \sum_i V(\Lambda_i)}.
\eeq
\textbf{Remark:} this can be done only if $t>0$. Otherwise this shouldn't really be called a matrix model.

\item the $\hbar$ expansion coefficients of correlation functions of this matrix model are given by the Topological Recursion invariants (modulo even terms). The moments are
\bea
\mathbb E_{\text{cumulant}}\left( \Tr M^{k_1} \dots \Tr M^{k_n}\right)
&=& \sum_{g=0}^\infty \hbar^{2g-2+n} \frac{1}{(2\pi\ii)^n} \cr 
&& \!\! \oint_{\mathcal C_+}\dots\oint_{\mathcal C_+}\  \omega_{g,n}(z_1,\dots,z_n)  \  \x(z_1)^{k_1}  \dots \x(z_n)^{k_n}  .\cr
\eea

\end{itemize}

%================================================
%================================================

\section{Conclusion}

Given a spectral curve on a cylinder with 2 branchpoints, we have reconstructed a formal random matrix measure, whose moments are given by topological expansion of the spectral curve.

A key step was to use the notion of generalized cycles introduced in \cite{EynGenCycles}.
It is only through them that we can get the 2-body interaction from the $B$.

The method could be generalized to identify more than 1-cut situations, and more generally, if a spectral curve has a geometry similar to that of a given well-understood model solvable by topological recursion, we can try to recover the parameters of the model from the spectral curve.

The geometry of a cylinder is a very useful one.
For example it is useful to understand the so-called "conifold singularity", i.e. what happens when a cylinder gets pinched to a nodal point. See \cite{brini2026conifoldgapallgenusmirror,fang2026conifoldgaptheoremtopological}.
We will also use it in a shortly coming work on gluing spectral curves.

%================================================
%================================================

\section*{Acknowledgements and Declarations}

\begin{itemize}
\item \textbf{Funding:} The author is employed by IPHT/DRF/CEA, and has not received external funding for this work.

    \item \textbf{AI use:} AI is now unavoidable.
Several computations were tested with GPT-Astra 6.2.
It was also used to polish the presentation.
\item \textbf {Competing interests:}
The author declares to have no competing interests.
\item \textbf {Data and code availability:}
No datasets were generated or analysed in this study.

\end{itemize}

\printbibliography

\end{document}